\documentclass{article}

\usepackage{arxiv}

\usepackage{cite}
\usepackage{amsmath,amssymb,amsfonts}
\usepackage{algorithm}
\usepackage{algorithmic}
\usepackage{graphicx}
\usepackage{textcomp}
\usepackage[table]{xcolor}
\usepackage[hyphens]{url}
\usepackage{fancyhdr}
\usepackage{booktabs}
\usepackage{pifont}
\usepackage{hyperref}
\usepackage{xspace}
\hypersetup{hidelinks}
\usepackage{multirow}
\usepackage{multicol}

\newcommand{\figref}[1]{\figurename~{\ref{#1}}}
\newcommand{\tabref}[1]{Table~{\ref{#1}}}

\newcommand{\eqnref}[1]{Eqn.~{(\ref{#1})}}

\newcommand{\yesicon}{\textcolor{green!45!black}{\ding{51}}}
\newcommand{\noicon}{\textcolor{red!70!black}{\ding{55}}}

\definecolor{salutbg}{RGB}{230,246,238}
\definecolor{salutFixed}{RGB}{232,247,235}
\definecolor{salutConfig}{RGB}{224,240,255}
\definecolor{salutDerived}{RGB}{255,245,214}
\definecolor{salutPolicy}{RGB}{242,232,255}
\definecolor{salutWorkload}{RGB}{235,243,250}
\definecolor{salutArchitecture}{RGB}{235,247,238}
\definecolor{salutSearch}{RGB}{245,239,250}

\title{Unified Lookup-Table Inference with Signed-Digit K/V Caches for Ternary LLMs}

\author{%
  \textbf{Ziang DUAN$^{\dagger\ddagger 1,2}$, Jiajun WU$^{\dagger 3,4}$, Zetian CHEN$^2$, Hao SONG$^2$, Yanwen DENG$^2$, Zixuan SHEN$^1$} \\
  \textbf{Nuobei XIE$^2$, Simo WU$^2$, Bolun WANG$^2$, Peng ZHOU$^2$, and Chao WANG$^1$} \\[0.2cm]
  \normalfont $^1$School of Optical and Electronic Information, Huazhong University of Science and Technology, China\\
  \normalfont $^2$LuxiTech Co. Ltd., Shenzhen, China\\
  \normalfont $^3$AI Chip Center for Emerging Smart Systems, Hong Kong SAR\\
  \normalfont $^4$Hong Kong University of Science and Technology, Hong Kong SAR \\[0.2cm]
  \small\texttt{duanzang@hust.edu.cn, jiajunwu@ust.hk, z773366@outlook.com, synapse787fx@gmail.com}\\
  \small\texttt{yan@luxitech.cn, d202281028@hust.edu.cn, roger@luxitech.cn, simo@luxitech.cn}\\
  \small\texttt{wangbolun@luxitech.cn, zp@luxitech.cn, chao\_wang\_me@hust.edu.cn}
}

\begin{document}
\maketitle

\def\thefootnote{$\dagger$}
\footnotetext{Ziang DUAN and Jiajun WU contributed equally to this work.}
\def\thefootnote{\arabic{footnote}}

\def\thefootnote{$\ddagger$}
\footnotetext{This work was conducted during an internship at LuxiTech Co. Ltd.}
\def\thefootnote{\arabic{footnote}}
\begin{abstract}
Ternary LLMs make their weight-dominated projections compact and efficient,
but attention remains a mismatch: its K/V cache is created online and is
typically processed by a separate higher-precision engine. Compressing this
cache alone does not resolve the mismatch. To execute attention with the same
lookup-table machinery as ternary projections, values accumulated in one
reduction must retain a compatible representation and scale. This requirement
also differs for keys and values during causal decoding, because newly
generated values may belong to an unfinished cache block.

This work develops a unified lookup-table inference approach for ternary LLMs.
It stores runtime K/V states as scaled multi-plane signed digits organized
around the reduction structure of attention. The resulting digit planes are
consumed directly by activation-derived tables, avoiding dense K/V
materialization between cache storage and attention computation. The design
combines online K/V formation, bounded handling of incomplete value blocks,
and a shared multi-stream datapath for Linear projections and attention. A
constraint-guided search selects the representation and execution policy for
a target quality--efficiency trade-off. Experiments on native and
post-training ternary models validate the approach across cache capacity,
model quality, and hardware efficiency.
\end{abstract}
\section{Introduction}
\label{sec:introduction}

\begin{figure}
    \centering
    \includegraphics[width=0.46\columnwidth]{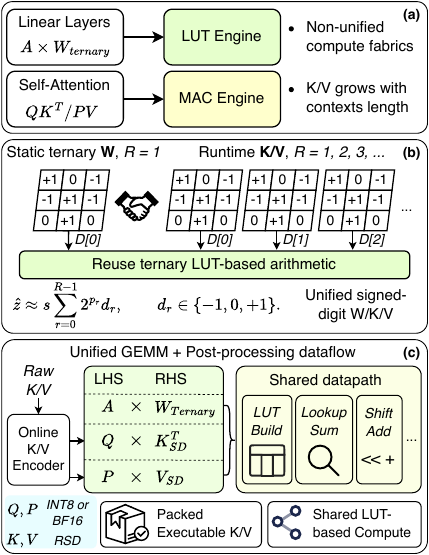}
    \caption{Signed-digit K/V representation and unified execution model.}
    \label{fig:intro-unified}
\end{figure}

Large language models (LLMs) support interactive applications such as
chatbots and code generation \cite{InstructGPT,GPT4,Codex,CodeLlama}, but
autoregressive inference remains costly because of repeated weight access and
the growing K/V cache \cite{DataMovement}. Model compression is therefore
crucial \cite{zhu2024survey}. Ternary LLMs
\cite{BitNet,BitNet158,BitNet2B4T,Spectra} constrain Linear weights to
$\{-1,0,+1\}$, reducing weight storage while enabling faster computation than
higher-precision models \cite{lut-survey}.

Grouped lookup-table (LUT) execution
\cite{UNPU,TernaryLUTGen,TMAC,LUTTensorCore,TellMe,TENET,LUTLLM,TMAN} is a
natural match for ternary Linear projections. It replaces a group of
weight--activation products with table construction, lookup, and reduction:
a fixed ternary weight group selects one activation-derived sum. The weight
addresses can be prepared offline and reused across tokens, while each runtime
table serves multiple output channels. This execution model has consequently
been realized across the stack, from LUT-table generators and processing
elements \cite{UNPU,TernaryLUTGen,TENET}, to optimized CPU kernels
\cite{TMAC}, and full LLM accelerators
\cite{LUTTensorCore,TellMe,LUTLLM,TMAN}.

However, existing ternary LLM accelerators commonly apply LUT execution only
to static Linear weights, while retaining separate integer or floating-point
datapaths for query--key multiplication and probability--value accumulation
($QK^\top$ and $PV$) in self-attention
\cite{TellMe,TENET,VitaLLM,LUTLLM}, as illustrated in
\figref{fig:intro-unified}(a). Some designs leave the K/V cache unquantized
and execute attention on floating-point K/V \cite{TellMe,LUTLLM}. Others
reduce K/V-cache capacity and bandwidth with low-bit codes, specialized
scaling, outlier handling, or residual windows, but materialize intermediate
floating-point K/V values for attention computation
\cite{KIVI,KVQuant,SKVQ,CommVQ,KVTuner,RateQuant}. In both cases, attention
cannot directly use the packed ternary-address format consumed by the LUT
fabric, requiring a separate compute path.

Some algorithmic methods retain ternary K/V representations during attention.
For example, KVTQ uses ternary residual channels to replace multiplications
with additions \cite{KVTQ}. Yet ternary digits alone do not make attention
directly executable on a Linear LUT engine. Values contributing to one LUT
reduction must share a scale; otherwise, each partial result requires
floating-point scaling and accumulation. This condition affects K and V
differently. In $QK^\top$, one K token produces one completed score, allowing
a token-local scale. In $PV$, each output value combines V data from many
tokens, so token-local V scales occur within the same sum. V must instead be
organized by value channel over a token interval, whose newest suffix remains
incomplete during causal decoding. These representation, layout, and
finalization constraints make ternary attention nontrivial to migrate onto a
Linear LUT engine.

To meet these requirements, we propose a quantization--architecture co-design
that extends directly executable LUT computation from static ternary weights
to dynamic K/V attention. \figref{fig:intro-unified}(b) illustrates the key
representation. Our system encodes each K/V block with one shared scale and
$R$ restricted signed-digit (RSD) planes at template-selected power-of-two
positions. Each plane contains ternary digits stored as packed LUT addresses;
plane weighting and scale recovery occur after lookup and reduction. Static
ternary weights are the $R=1$ case, whereas runtime K/V uses additional
planes when needed without reconstructing dense values before attention. This
common representation lets Linear layers, $QK^\top$, and $PV$ use the same
LUT interface. During autoregressive decoding, K blocks are finalized when
their tokens are produced, while an incomplete V interval remains in a bounded
tail until it forms a scale-consistent packed block.

\figref{fig:intro-unified}(c) shows how the proposed system turns the
representation and layout requirements into one execution path. Its
cache-resident RSD format encodes dynamic K/V as packed ternary addresses with
a shared block scale, while the operator-aligned K/V layout preserves causal
cache updates. An online encoder produces these addresses at runtime; Linear
layers, $QK^\top$, and $PV$ then share table construction, lookup, and
post-processing without reconstructing finalized dense K/V. Concurrent
streams co-schedule blocks with the same plane count to maintain compatible
execution latency. The A8 configuration constructs LUTs from dynamically
quantized INT8 activation-side operands, whereas the A16 configuration retains
BF16 operands; both use the same RSD K/V representation.

The representation, block geometry, plane count, template gaps, and stream
mapping jointly determine quality, capacity, and execution time. We therefore
use a hardware-aware DSE to select legal representation and mapping policies
under a quality constraint. Across native and QAT-converted ternary models,
the selected policies reduce packed K/V payload to as little as $21\%$ of
BF16. Relative to throughput-matched heterogeneous baselines, the A8
configuration improves attention throughput per area and per power by
$2.52\times$ and $2.30\times$ on Falcon-E-3B, while the A16 configuration
improves them by $4.78\times$ and $8.61\times$ on OPT-350M, respectively.
\section{Background and Related Work}
\label{sec:background}

\subsection{Ternary Models and LUT Compute}

BitNet b1.58 2B4T and Spectra use ternary weights in $\{-1,0,+1\}$ for the dominant linear layers \cite{BitNet2B4T,Spectra}. Such models can be trained from scratch with quantization-aware training (QAT), and QAT can also convert pretrained full-precision LLMs into ternary variants \cite{TernaryLLM,Hestia,Sherry}. The low precision of ternary weights makes LUT computation a natural alternative to fixed-point MAC: a ternary group indexes a finite set of activation sums. Compared with fixed-point MAC arrays, LUT execution replaces per-product multiplication with table construction, lookup, and reduction, and its energy advantage grows as operand precision decreases \cite{UNPU}. T-MAC realizes grouped lookup on CPUs, LUT Tensor Core maps the principle to a custom tensor-core organization, and recent generators explore specialized LUT datapaths \cite{TMAC,LUTTensorCore,TernaryLUTGen}. T-MAN applies table lookup to low-bit operations that are inefficient on mobile NPUs \cite{TMAN}.

At the system level, ternary accelerators apply LUT-based computation to linear projections and execute attention through separate integer- or BF16-based datapaths. TeLLMe combines table-lookup ternary linear with dedicated attention logic \cite{TellMe}. TENET couples sparse ternary LUT cores with high-precision cores \cite{TENET}, while VitaLLM schedules ternary projections and INT8 attention across distinct arithmetic resources \cite{VitaLLM}. LUT-LLM uses two-dimensional lookup for linear and a separate attention dataflow \cite{LUTLLM}. These systems retain distinct arithmetic paths for ternary linear layers and attention. The proposed system provides a common LUT substrate for both operations.

\subsection{K/V Cache Quantization}

K/V cache size and bandwidth scale linearly with sequence length. KIVI leverages asymmetric statistics via per-channel K, per-token V, and a full-precision residual window \cite{KIVI}. KVQuant, SKVQ, and QAQ introduce outlier handling, smoothing, sliding windows, and adaptive precision \cite{KVQuant, SKVQ, QAQ}. CommVQ learns additive codebooks compatible with rotary embeddings, while AQUA-KV and Oaken fuse offline analysis with online cache encoding \cite{CommVQ,AQUAKV,Oaken}. KVTQ encodes K/V as multiple ternary residual digits \cite{KVTQ}.
Specifically, KVTuner assigns a precision level to each layer,
RateQuant allocates bits across heads using calibrated curves of
rate versus distortion, InnerQ incorporates hardware cost into
K/V cache quantization, and Titanus jointly models dependent cache
tiers during online pruning and quantization
\cite{KVTuner, RateQuant, InnerQ, Titanus}.
Several prior methods also defer V quantization until enough tokens
have accumulated to form a complete block
\cite{KIVI, HFQuantizedCache}.
The proposed system encodes the resulting K/V blocks as packed RSD planes with
a shared scale, making dynamic K/V directly executable on the LUT substrate
used for static ternary weights.

\subsection{Hardware-Aware Quantization}

Hardware-aware LLM accelerators exploit low-bit or adaptable formats,
bit-serial execution, and LUT arithmetic
\cite{zadeh2022mokey,wang2022via,AutoNBA,ANT,OliVe,MSD,BitMoD,VitaLLM,
TellMe,LUTTensorCore}, as summarized in
\tabref{tab:qualitative-comparison}. Mixed precision further exploits
nonuniform quantization sensitivity to balance accuracy and hardware cost
\cite{HAQ,ANT,MSD,HAWQ,HAO,DyBit}. The proposed system extends this principle
by jointly selecting signed-digit K/V policies and legal mappings for a
unified LUT fabric.

\begin{table}[tbp]
  \centering
  \caption{Qualitative Comparison with Representative Related Works}
  \label{tab:qualitative-comparison}
  \small
  \setlength{\tabcolsep}{1.25pt}
  \renewcommand{\arraystretch}{1.03}
  \begin{tabular}{@{}lcccc@{}}
    \toprule
    \multirow{2}{*}{\textbf{Work}} & \textbf{Compute} & \textbf{Applied} & \textbf{Mixed} & \textbf{HW-aware} \\
                                  & \textbf{fabric}  & \textbf{format}  & \textbf{precision} & \textbf{DSE} \\
    \midrule
    Auto-NBA~\cite{AutoNBA} & MAC & INT & \yesicon & \yesicon \\
    ANT~\cite{ANT} & MAC & INT/FLINT/PoT & \yesicon & \noicon \\
    OliVe~\cite{OliVe} & MAC & INT/FLINT/Abfloat & \yesicon & \noicon \\
    MSD~\cite{MSD} & Bit-serial & INT/SD & \yesicon & \yesicon \\
    BitMoD~\cite{BitMoD} & Bit-serial & INT/FP & \noicon & \noicon \\
    VitaLLM~\cite{VitaLLM} & Fused MAC & INT/ternary & \yesicon & \noicon \\
    TeLLMe~\cite{TellMe} & LUT & INT/ternary & \noicon & \noicon \\
    LUT Tensor Core~\cite{LUTTensorCore} & LUT & INT/FP & \yesicon & \yesicon \\
    \rowcolor[rgb]{0.902,0.965,0.933}
    \textbf{This Work} & \textbf{LUT} & \textbf{BF16/INT/SD} & \yesicon & \yesicon \\
    \bottomrule
  \end{tabular}
\end{table}

\section{Arithmetic and Quantization}
\label{sec:arithmetic}

This section first formulates linear and attention products under a common
GEMM interface, and then extends ternary LUT arithmetic to runtime K/V
through a block-wise restricted signed-digit representation.

\subsection{LLM Inference and Ternary GEMM}

For layer $\ell$, the input activations are projected into queries, keys, and
values:
\begin{equation}
  \begin{aligned}
    [\mathbf Q^\ell\ \mathbf K^\ell\ \mathbf V^\ell]
    &=
    \mathbf X_{\mathrm{in}}^\ell
    [\mathbf W_Q^\ell\ \mathbf W_K^\ell\ \mathbf W_V^\ell],
    \\
    \mathbf X_{\mathrm{out}}^\ell
    &=
    \operatorname{MLP}(\mathbf A^\ell).
  \end{aligned}
  \label{eq:prefill-linear-simplified}
\end{equation}
For head $h$, let $M$ and $N$ denote the query and K/V sequence lengths,
respectively, where $M=N$ during prefill and $M=1$ during decode. With
$d_k=d/n_h$, attention is computed as
\begin{equation}
  \begin{aligned}
    \mathbf P^{\ell,h}
    &=
    \operatorname{softmax}\left(
      \frac{\mathbf Q^{\ell,h}(\mathbf K^{\ell,h})^\top}{\sqrt{d_k}}
    \right),
    \\
    \mathbf A^{\ell,h}
    &=
    \mathbf P^{\ell,h}\mathbf V^{\ell,h}.
  \end{aligned}
  \label{eq:prefill-attention-simplified}
\end{equation}
where
$\mathbf Q^{\ell,h}\in\mathbb R^{M\times d_k}$,
$\mathbf K^{\ell,h},\mathbf V^{\ell,h}\in\mathbb R^{N\times d_k}$,
$\mathbf P^{\ell,h}\in\mathbb R^{M\times N}$, and
$\mathbf A^{\ell,h}\in\mathbb R^{M\times d_k}$.

We express all matrix products using the operator-local convention
\begin{equation}
  \mathbf Y[M,N]
  =
  \mathbf X[M,K]\mathbf Z[K,N],
  \label{eq:gemm-convention}
\end{equation}
where $\mathbf X$ and $\mathbf Z$ are the left-hand-side (LHS) and
right-hand-side (RHS) operands. Activations, queries, and attention
probabilities use the LHS path, while static weights, cached keys, and cached
values use the RHS path, enabling linear, $\mathbf Q\mathbf K^\top$, and
$\mathbf P\mathbf V$ to share one LUT interface.

For one output entry, LUT-based ternary GEMM partitions the RHS operand into
groups of $G$ ternary values and precomputes all $3^G$ possible signed sums
of the corresponding LHS activations \cite{TellMe,TernaryLUTGen}. Let
$\mathcal A_G=\{-1,0,+1\}^{G}$. A ternary group
$\mathbf a\in\mathcal A_G$ is mapped to the LUT address
\begin{equation}
  \operatorname{idx}_G(\mathbf a)
  =
  \sum_{j=0}^{G-1}(a_j+1)3^{G-1-j}.
  \label{eq:lut-address}
\end{equation}
For $G=3$, the $3^3=27$ ternary states fit in a five-bit binary address. For
example, $\mathbf a=(-1,0,+1)$ maps to index $5$, represented as
$\texttt{00101}$.

For LUT group $g$, the precompute unit constructs
\begin{equation}
  \mathcal T_g[\operatorname{idx}_G(\mathbf a)]
  =
  \sum_{j=0}^{G-1} X_{(g \times G + j)} a_j.
  \label{eq:lut-table}
\end{equation}
For ternary RHS block $b$, let $\mathbf w_g$ denote its $g$-th ternary group
and $s_b$ its block scale. The output is obtained by reducing the selected
LUT entries over groups and blocks:
\begin{equation}
  \widehat{Y}
  =
  \sum_b s_b
  \sum_g
  \mathcal T_g\!\left[
    \operatorname{idx}_G(\mathbf w_g)
  \right].
  \label{eq:lut-ternary-gemm}
\end{equation}
Equation~\eqref{eq:lut-ternary-gemm} defines the ternary linear LUT
interface, which we next extend to runtime K/V without dense reconstruction.

\subsection{Block Signed-Digit Quantization}

\begin{figure}[tbp]
    \centering
    \includegraphics[width=0.56\columnwidth]
    {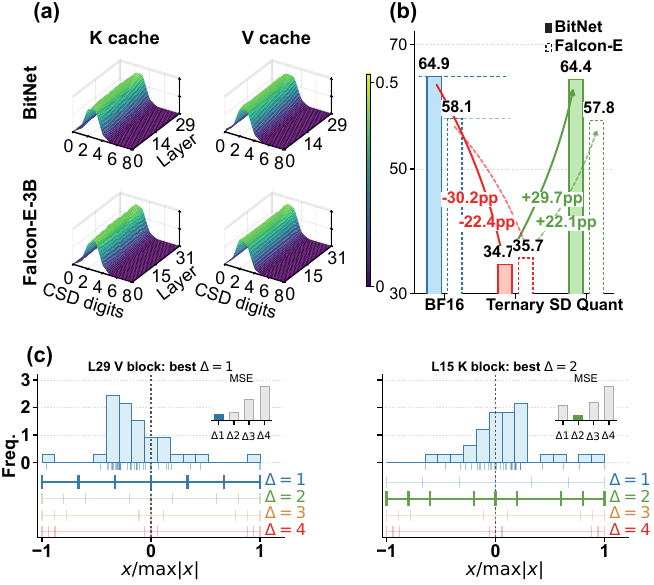}
    \caption{Signed-digit K/V quantization analysis. (a) Nonzero-digit
    distribution of canonical-CSD INT8 K/V in BitNet and Falcon-E-3B.
    (b) Multiple-choice accuracy with BF16, ternary, and three-plane K/V.
    (c) Normalized block distributions, two-plane codebooks, and
    corresponding quantization MSEs for representative K and V blocks.}
    \label{fig:kv-csd-motivation}
\end{figure}

\figref{fig:kv-csd-motivation}(a) shows that canonical-CSD INT8 K/V values
in BitNet and Falcon-E-3B typically contain only two or three nonzero digits,
motivating a compact multi-plane representation \cite{MSD}. As shown in
\figref{fig:kv-csd-motivation}(b), a single ternary plane causes substantial
accuracy degradation, whereas three planes largely recover BF16 accuracy.
The proposed system therefore represents each K/V block $b$ with $R_b$ RSD
planes. Static ternary weights are the special case $R=1$ and retain the LUT
interface in \eqnref{eq:lut-ternary-gemm}.

For a BF16 K/V block $Z^{(b)}$, let
$\mathbf p_b=\{p_b[0],\ldots,p_b[R_b-1]\}$ denote the integer bit positions
assigned to its $R_b$ signed-digit planes. Plane $r$ therefore contributes
$d_r2^{p_b[r]}$, with larger positions carrying greater significance.
The position set determines the normalized codebook, while a shared scale
sets the block magnitude:
\begin{equation}
  q_b
  =
  \sum_{r=0}^{R_b-1}2^{p_b[r]},
  \qquad
  s_b
  =
  \frac{\max_i|Z_i^{(b)}|}{q_b}.
  \label{eq:template-normalization}
\end{equation}
The corresponding normalized codebook is
\begin{equation}
  \mathcal C_b
  =
  \left\{
    \sum_{r=0}^{R_b-1}d_r2^{p_b[r]}
    \;\middle|\;
    d_r\in\{-1,0,+1\}
  \right\}.
  \label{eq:template-codebook}
\end{equation}
Each normalized K/V value is mapped to its nearest codeword. Midpoint ties
select the even-ranked codeword, while duplicate digit tuples use the
lexicographically first tuple.

For an LHS slice $\mathbf X_g$ and the $r$-th RSD digit group
$\mathbf d_{g,r}$, substituting the signed-digit representation into
\eqnref{eq:lut-table} gives
\begin{equation}
  \begin{aligned}
    \widehat{Y}
    &=
    \sum_b\sum_g
    \mathbf X_g^\top
    \left(
      s_b\sum_{r=0}^{R_b-1}
      2^{p_b[r]}\mathbf d_{g,r}
    \right) \\
    &=
    \sum_b s_b
    \sum_{r=0}^{R_b-1}2^{p_b[r]}
    \sum_g
    \mathcal T_g\!\left[
      \operatorname{idx}_G(\mathbf d_{g,r})
    \right].
  \end{aligned}
  \label{eq:lut-signed-digit-gemm}
\end{equation}
Each RSD plane reuses the ternary LUT addressing rule, and the plane results
are shifted and accumulated after lookup. For $R_b=1$ and
$\mathbf d_{g,0}=\mathbf w_g$,
\eqnref{eq:lut-signed-digit-gemm} reduces to
\eqnref{eq:lut-ternary-gemm}, unifying ternary weights and RSD K/V under one
executable LUT representation.

A common exponent shift can be absorbed into the block scale. For
$p'_b[r]=p_b[r]+\kappa$,
\begin{equation}
  q'_b=2^\kappa q_b,
  \qquad
  s'_b=2^{-\kappa}s_b,
  \label{eq:scale-shift}
\end{equation}
and
\begin{equation}
  s'_b\sum_r d_r2^{p'_b[r]}
  =
  s_b\sum_r d_r2^{p_b[r]}.
  \label{eq:scale-shift-equivalence}
\end{equation}
Thus, a common shift preserves both reconstruction and packed encoding.
The proposed system stores only the relative plane configuration
\begin{equation}
  \tau_b=(R_b,\boldsymbol{\Delta}_b),
  \label{eq:rsd-template}
\end{equation}
where $R_b$ is the plane count and $\boldsymbol{\Delta}_b$ records the
exponent gaps. The first plane is fixed at zero:
\begin{equation}
  p_b[0]=0,
  \qquad
  p_b[r]=\sum_{j=0}^{r-1}\Delta_b[j].
  \label{eq:template-positions}
\end{equation}

At fixed $R_b=2$, \figref{fig:kv-csd-motivation}(c) shows that
$\Delta=1$ minimizes the MSE for the representative V block, while
$\Delta=2$ is preferred for the K block. Adjusting the exponent gaps adapts
the reconstruction levels to each block distribution, improving fidelity
without increasing the plane count or execution cycles.

With $G=3$ packing, the digit-plane storage is
\begin{equation}
  B_{\mathrm{digit}}
  =
  5\sum_b R_b
  \left\lceil\frac{n_b}{3}\right\rceil
  \quad\text{bits},
  \label{eq:digit-storage}
\end{equation}
where $n_b$ is the number of valid values in block $b$; scales and template
identifiers are stored separately. Hence, $R_b$ determines K/V cache
footprint and plane-loop cost, while $\boldsymbol{\Delta}_b$ adjusts fidelity
at fixed $R_b$. This separation motivates the mixed-$(R,\boldsymbol{\Delta})$
optimization in Section~\ref{sec:dse}.

\subsection{Executable K/V Layout and Causal Finalization}
\label{sec:executable-kv-layout}

The factorization in \eqnref{eq:lut-signed-digit-gemm} imposes an execution
constraint. For one RHS output column and reduction block $b$, direct LUT
execution requires
\begin{equation}
  \mathbf X^{(b)}\widehat{\mathbf z}^{(b)}
  =
  s_b\sum_{r=0}^{R_b-1}2^{p_b[r]}
  \mathbf X^{(b)}\mathbf d_r^{(b)},
  \label{eq:executable-rsd-block}
\end{equation}
where $\mathbf d_r^{(b)}$ is the $r$-th ternary RHS plane. All entries in
$b$ must share $s_b$, allowing rescaling after LUT lookup and plane
reduction; otherwise, per-subblock rescaling would break the unified
post-lookup schedule.

The two attention products require different RHS layouts:
\begin{equation}
  \begin{aligned}
    \mathbf Q[M,d_k]\mathbf K^\top[d_k,N]
    &\rightarrow \mathbf P[M,N],\\
    \mathbf P[M,N]\mathbf V[N,d_k]
    &\rightarrow \mathbf A[M,d_k].
  \end{aligned}
  \label{eq:attention-rhs-layout}
\end{equation}
For $\mathbf Q\mathbf K^\top$, RHS columns are cached tokens and reduction
traverses $d_k$, so K blocks are token-local and finalized when each key is
produced. For $\mathbf P\mathbf V$, RHS columns are value channels and
reduction traverses the causal token dimension, so V blocks span token
intervals. Before deployment, the DSE statically scans legal templates for
each operator/layer/head/channel-block class and fixes
$\tau_b=(R_b,\boldsymbol{\Delta}_b)$. All runtime instances reuse this
template.

Appending a value to an unfinished V interval may change its maximum and
scale, invalidating earlier RSD codes. The proposed system therefore keeps one
bounded V-tail of at most $B_V$ tokens, containing authoritative input-precision
values and two packed RSD banks. After update $t$, it recomputes
$s_{b,t}$ over the $\ell_t$ valid entries and re-encodes the complete prefix
with the fixed $\tau_b$ into the inactive bank. Once encoding completes, it
atomically commits the bank selector with $(\ell_t,s_{b,t})$ and permits
$\mathbf P\mathbf V$ only after both softmax and the commit complete.
Consequently, $\mathbf P\mathbf V$ never observes an incomplete prefix or
codes generated under different scales.

Tail encoding overlaps with $\mathbf Q\mathbf K^\top$ and softmax. The
latency model charges $\ell_t d_k$ encoding operations and the corresponding
tail-buffer accesses before applying this overlap. The work is bounded by
$B_Vd_k$ values per head and is independent of context length. When
$\ell_t=B_V$, the active bank is committed directly as a finalized packed
block and the tail is reused. Thus, per request and layer, dense V state is
bounded by one open interval, and no finalized block is rewritten.
\begin{figure*}[t]
  \centering
  \includegraphics[width=\linewidth]{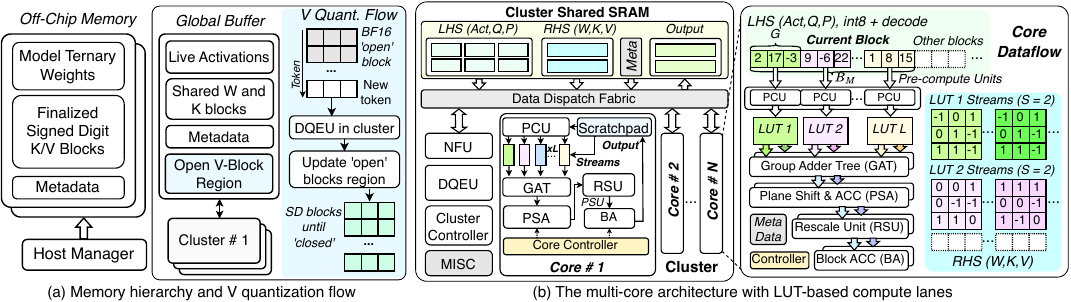}
  \caption{Overview of the proposed system. (a) Memory hierarchy and online
  V-block quantization with an Open V-Block Region. (b) Multicore cluster and
  core dataflow with independent RHS streams through shared PCU/LUT pipelines.}
  \label{fig:arch-diag}
\end{figure*}

\section{Hardware Architecture and Dataflow}
\label{sec:hardware}

\begin{figure}[tbp]
  \centering
  \includegraphics[width=0.58\columnwidth]{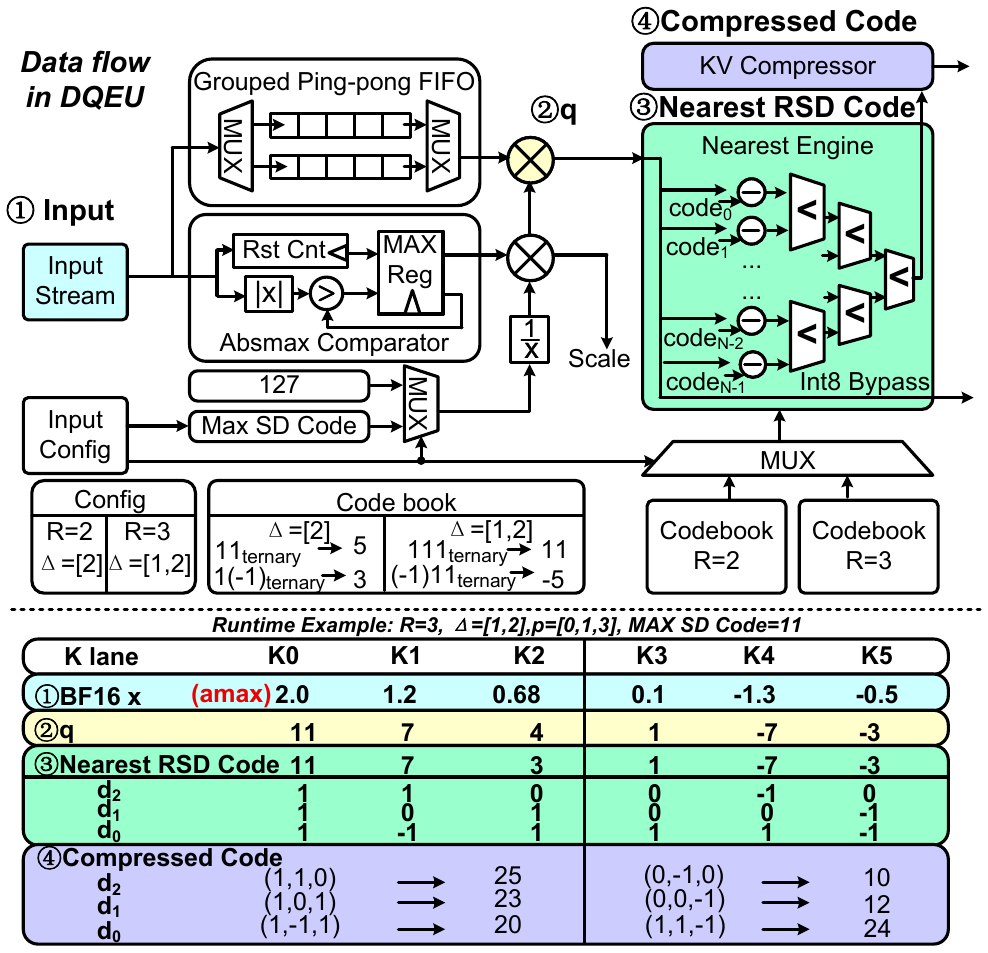}
  \caption{DQEU datapath, shown for the A8 configuration. The A16
  configuration bypasses the activation branch and reuses the K/V encoding
  branch.}
  \label{fig:dqeu}
\end{figure}

The proposed system couples the signed-digit representation with a unified
multicore LUT architecture for linear, $QK^\top$, and $PV$. This section
presents its core organization, online encoding, tiled streaming dataflow, and
operator latency model.

\subsection{Hardware System Design}

\figref{fig:arch-diag}(a) shows the memory hierarchy and V-block lifecycle.
Off-Chip Memory stores Model Ternary Weights, Finalized Signed Digit K/V
Blocks, and metadata. The Global Buffer (GB) holds Live Activations, Shared W
and K Blocks, metadata, and the bounded V-tail in its Open V-Block Region. New
V tokens enter through the cluster DQEU, and finalized blocks are committed to
Off-Chip Memory.

\figref{fig:arch-diag}(b) shows the cluster organization. The Data Dispatch
Fabric distributes LHS $(A/Q/P)$, RHS $(W/K/V)$, metadata, and output tiles
between Cluster Shared SRAM, the compute cores, and the shared NFU and DQEU.
The nonlinear function unit (NFU) handles masking, attention scaling, and
softmax, while the Cluster Controller and miscellaneous control logic (MISC)
enforce tile-level dependencies.

Each core contains a Scratchpad, a Core Controller, $L$ paired Pre-Compute
Units (PCUs) and LUTs, and up to $S_{\max}$ independent RHS streams. Each PCU
builds the $3^G$ LUT entries, which the streams access through independent
read ports. For example, with $B_K=32$ and $G=3$, $L=11$ LUT units cover 33
positions with one padding position. Two streams process two output columns
concurrently, providing 64 useful RHS values per plane cycle. Within each
stream, the Group Adder Tree (GAT) sums LUT outputs, the Plane Shifter \&
Accumulator (PSA) combines the signed-digit planes, the Rescale Unit (RSU)
applies the block scale, and the Block Accumulator (BA) retains partial sums
until the full reduction completes.

The core dataflow in \figref{fig:arch-diag}(b) illustrates the A8
configuration, whose INT8 $A/Q/P$ values build the LUTs before the RSU applies
activation and RHS scales. The A16 configuration shares this dataflow, but
BF16 $A/Q/P$ bypass activation quantization and build BF16 table entries, so
its DQEU encodes only K/V.

The Open V-Block Region stores the authoritative BF16 V-tail block and two
signed-digit shadow banks. On each append, the DQEU updates the inactive
shadow while attention reads the active one, after which the controller
atomically swaps them. Once complete, the active shadow and its metadata are
committed as a Finalized Signed Digit V Block in Off-Chip Memory, and the
region is reused. Token-local K is encoded and appended directly, so finalized
cache blocks are never rewritten.

\subsection{Dynamic Quantization and Encoding Unit}

\figref{fig:dqeu} shows the Dynamic Quantization and Encoding Unit (DQEU).
A grouped ping-pong FIFO overlaps block capture and encoding. The
absmax/scaling path, Nearest Engine, and K/V Compressor derive block scales,
select RSD digits, and pack LUT addresses.

For block $b$, $\tau_b=(R_b,\boldsymbol{\Delta}_b)$ determines $q_b$ and
$\mathcal C_b$, while the MAX register provides
$\max_i|Z_i^{(b)}|$ to derive $s_b$ using
\eqnref{eq:template-normalization}. The Nearest Engine selects codewords, and
the compressor packs the resulting planes in $G=3$ groups using
\eqnref{eq:lut-address}. The lower example uses $\tau_b=(3,[1,2])$, giving
positions $[0,1,3]$ and $q_b=11$. For maximum $2.0$, $s_b=2.0/11$: values
$2.0$ and $-1.3$ map to $11$ and $-7$, represented by
$(d_2,d_1,d_0)=(1,1,1)$ and $(-1,0,1)$. The first three values produce plane
addresses $25$, $23$, and $20$.

For a V-tail update, the DQEU combines the retained input-precision prefix
with newly projected V values, recomputes the shared scale, and re-encodes
the complete prefix into the inactive bank while the committed bank remains
unchanged. One DQEU is provisioned to sustain the K/V production rate of one
core even during a full $R=3$ V-tail rebuild; token-local K encoding is less
demanding because it processes no retained prefix. The grouped ping-pong FIFO
decouples these operations and allows encoding to overlap with continued
Projection execution without creating a DQEU backlog. After encoding, the
bank selector, valid length, and scale are committed atomically before $PV$
accesses the updated tail. The A8 configuration also emits INT8
$\mathbf A/\mathbf Q/\mathbf P$, whereas the A16 configuration uses the DQEU
only for K/V.

\subsection{Tiled Streaming Dataflow and Scheduling}

\begin{figure}[tbp]
  \centering
  \includegraphics[width=0.6\columnwidth]{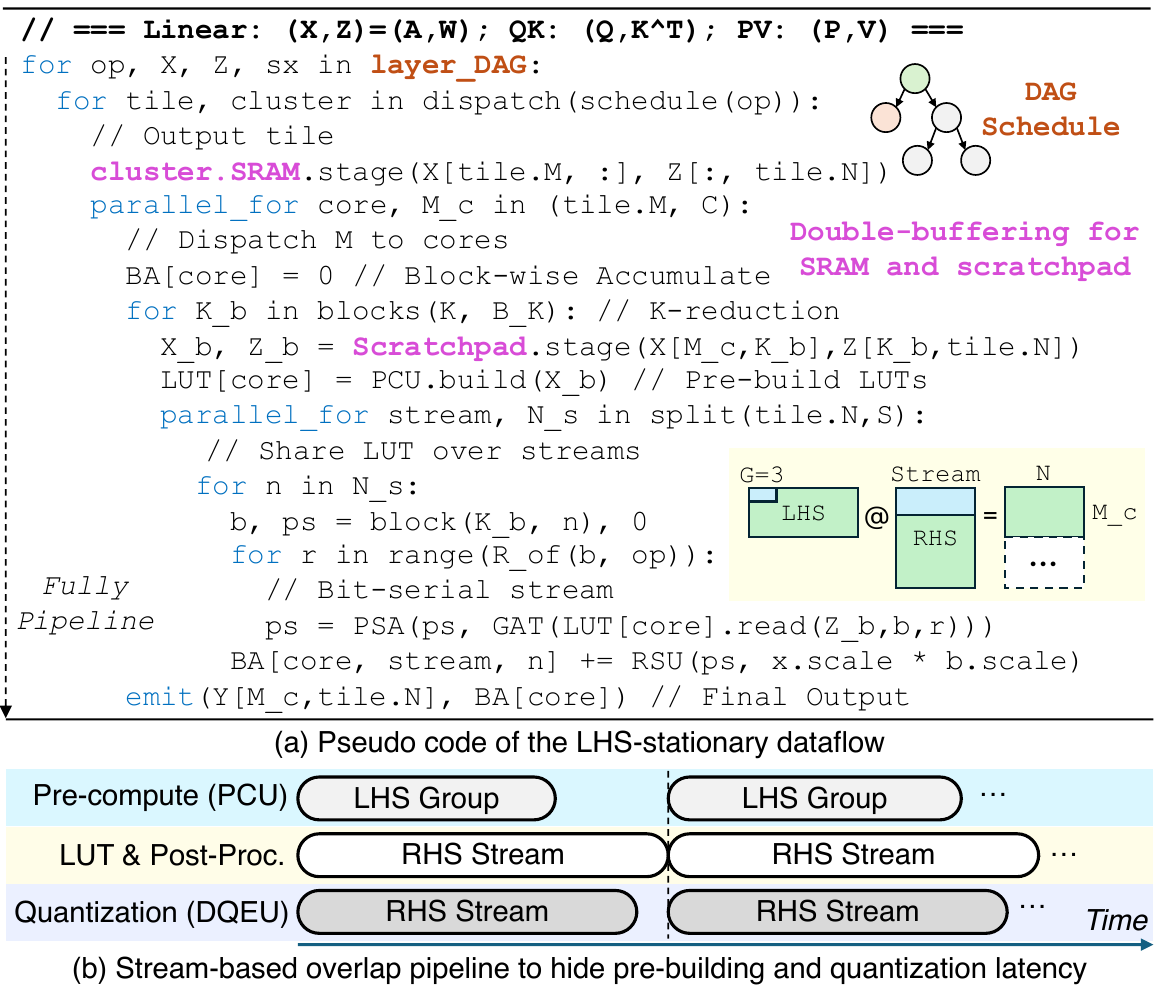}
  \caption{Tiled streaming dataflow. (a) The hierarchical LHS-stationary loop
  nest shared by linear, $QK^\top$, and $PV$. (b) Steady-state overlap
  pipeline.}
  \label{fig:dataflow-diag}
\end{figure}

\figref{fig:dataflow-diag}(a) unifies linear, $QK^\top$, and $PV$ as the same
matrix product $X^{M\times K}Z^{K\times N}$. Each $T_M\times T_N$ output
\emph{tile} maps to one cluster and carries the full $K$ reduction, avoiding
inter-tile accumulation. Within a cluster, cores shard $M$, and each core's
active \emph{streams} shard $N$. For each $B_K$-wide reduction \emph{block},
the scratchpad stages the LHS/RHS slices, and the PCU builds one LHS-derived
LUT image shared by all streams. Each stream reads an RHS quantization
\emph{block} over its $R$ planes, and BA commits only after the full
reduction. Only axis roles differ: $A/Q/P$ are activation-like LHS operands,
and $W/K^\top/V$ are RHS streams. $QK^\top$ streams cached-token columns and
reduces the head dimension, making K token-local; $PV$ streams V channels and
reduces context, so V is per-channel when $B_N^V=1$. The plane-loop bound $R$
is set by a policy-level $R$-group: QK groups tokens at fixed K/V head and
channel block, while PV groups channel blocks at fixed K/V head and token
block, with $\boldsymbol{\Delta}$ remaining block-wise.

\figref{fig:dataflow-diag}(b) shows the steady-state producer-consumer
overlap. While LUT and post-processing consume a ready RHS stream from one
table bank, the PCU builds the next LHS table in the alternate bank and the
DQEU prepares a future activation or K/V block in its ping-pong buffers.
Because these engines process different work on separate resources,
preparation is hidden whenever it finishes before the concurrent RHS; only
pipeline fill, drain, or a slower stage remains exposed. The same
dependency-ready rule lets Q/K/V projections, K/V encoding, $QK^\top$, the
NFU, and $PV$ overlap across tiles.

\subsection{Operator Latency Model}

Consider one GEMM $X^{M\times K}Z^{K\times N}$ under fixed hardware $H$ and
schedule $\pi=(T_M,T_N,B_K,C_{\rm act},S_{\rm act},\ldots)$. The schedule
creates $\lceil M/T_M\rceil\lceil N/T_N\rceil$ output tiles, each retaining
the full $K$ reduction across $N_B=\lceil K/B_K\rceil$ blocks. Within a tile,
$C_{\rm act}$ cores partition its $M$ rows and
$S_{\rm eff}=\min(S_{\rm act},S_{\max},N_t)$ streams per core partition its
$N_t$ valid columns.

For reduction block $b$, let $C_{\rm build,b}$ be its PCU table-build time and
$C_{\rm exec,b}$ the slowest active core/stream time through LUT lookup and
the pipelined GAT/PSA/RSU/BA stages, using each RHS block's group-constrained
$R$ and exact tails. Denote the operator latency by
$C_{\rm op}=C_{\rm op}(M,K,N;H,\pi,\mathbf R)$. Double-buffered LUT banks
overlap build of block $b{+}1$ with execution of block $b$, so the simulator
computes
\begin{equation}
  \begin{aligned}
    C_{\rm tile}={}&C_{\rm setup}+C_{\rm build,0}\\
    &+\sum_{b=0}^{N_B-2}\max\!\left(C_{\rm exec,b},C_{\rm build,b+1}\right)\\
    &+C_{\rm exec,N_B-1}+C_{\rm emit},\\
    C_{\rm op}={}&\operatorname{Schd}_{H,\pi}
    \!\left(\{C_{\rm tile}^{(i,j)}\}_{i,j}\right).
  \end{aligned}
  \label{eq:operator-latency}
\end{equation}
Here, $\operatorname{Schd}$ refers to the schedules of all output tiles on the
available clusters. The max term exposes the intended overlap: LUT
construction is hidden when stream execution dominates; $R$ changes plane
work, whereas $\boldsymbol\Delta$ at fixed $R$ is timing-invariant.
\section{Quantization-Dataflow Co-Optimization}
\label{sec:dse}

\subsection{Framework Overview}

\begin{figure}[tbp]
    \centering
    \includegraphics[width=0.54\columnwidth]{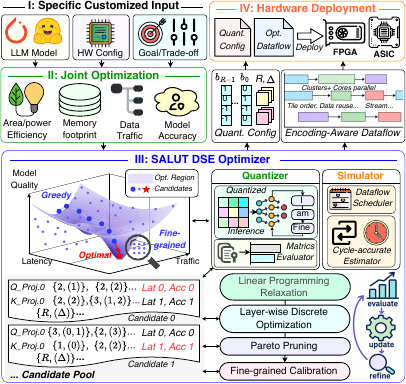}
    \caption{Multi-fidelity co-optimization flow.}
    \label{fig:dse-framework}
\end{figure}

K/V signed-digit precision and hardware mapping must be optimized jointly,
but exhaustive group-wise enumeration is intractable. The proposed system
therefore uses the four-stage multi-fidelity flow in
\figref{fig:dse-framework}.

Stage I specifies the pretrained model, target hardware, quantization
constraints, and deployment objective. Stage II constructs the joint search
space of signed-digit policies and legal dataflow mappings under fixed
area, power, capacity, and dependency constraints. Stage III uses analytical
surrogates to generate and prune candidates, the formulation-based simulator
to determine deployment cost, and held-out inference to validate quality.
Stage IV fixes the selected policy and its best legal mapping for FPGA or
ASIC deployment. The following
subsections formalize the policy, cost, and multi-fidelity search.

\subsection{Problem Definition}

Let $H$ denote the fixed accelerator and its $R$-group contract. A policy
$p=(\mathbf B,\{R_w\},\{\boldsymbol\Delta_b\})$ specifies the K/V block
shape $\mathbf B$, a plane count $R_w\in\{1,2,3\}$ for each RHS group $w$,
and an exponent-gap template $\boldsymbol\Delta_b$ for each block $b$.
Co-scheduled members share $R_w$ for compatible plane latency but retain
independent $\boldsymbol\Delta_b$. A mapping $\pi$ may change tiling, active
cores and streams, loop order, multicast, and cache residency, but not the
policy grouping. Let $\mathcal S(p)$ be the finite set of mappings satisfying
capacity, bandwidth, and dependency constraints.

For policy $p$ and mapping $\pi$, the simulator returns the lexicographically
ordered deployment cost
\begin{equation}
  C(H,p,\pi)=\bigl(L(H,p,\pi),\,T_{\rm ext}(H,p,\pi),\,M_{\rm GB}(H,p,\pi)\bigr),
  \label{eq:dse-deployment-cost}
\end{equation}
where $L$, $T_{\rm ext}$, and $M_{\rm GB}$ are whole-request latency,
off-chip traffic, and peak Global Buffer usage, respectively; the proposed
system minimizes them in this order. Let $\Delta\mathrm{NLL}(p)$ denote the
increase in held-out mean token-level negative log-likelihood (NLL) over dense
BF16 K/V, and let $\epsilon$ be the allowed quality-loss bound. From candidate
set $\mathcal C$, the proposed system selects
\begin{equation}
\begin{aligned}
  \pi^\star(p)&=\arg\min_{\pi\in\mathcal S(p)} C(H,p,\pi),\\
  \widehat p&=\arg\min_{p\in\mathcal C,\,\Delta\mathrm{NLL}(p)\leq\epsilon}
  C\!\left(H,p,\pi^\star(p)\right).
\end{aligned}
\label{eq:joint-dse}
\end{equation}
The inner search exhaustively evaluates the declared finite mapping set
$\mathcal S(p)$, making $\pi^\star(p)$ optimal within that set. The outer
search evaluates a compact candidate frontier $\mathcal C$. Since
$\mathrm{PPL}(p)/\mathrm{PPL}_{\rm BF16}
=\exp(\Delta\mathrm{NLL}(p))$, the NLL budget $\epsilon$ induces a
relative-perplexity bound and screens candidate policies before held-out
validation.

\subsection{R-Group-Aware Multi-Fidelity Search}

\begin{algorithm}[tbp]
  \caption{Multi-Fidelity Policy Search}
  \label{alg:multifidelity-dse}
  \small
  \begin{algorithmic}[1]
    \REQUIRE Model $f$, calibration and validation data, hardware $H$,
    legal block shapes $\mathfrak B$, NLL limit $\epsilon$
    \ENSURE Selected design $(\widehat p,\pi^\star(\widehat p))$
    \STATE $\mathcal C\leftarrow\varnothing$
    \FOR{each $\mathbf B\in\mathfrak B$ compatible with $H$}
      \STATE Select $t^\star_{b,r}$ and build $E_{w,r}$ and
      $\widehat C_{w,r}$
      \FOR{each excess-error budget $\tau$}
        \STATE Solve \eqnref{eq:dse-proxy-relaxation}
        \STATE Project and repair; add the resulting policy to $\mathcal C$
      \ENDFOR
    \ENDFOR
    \STATE Remove duplicate and surrogate-dominated policies
    \FOR{each $p\in\mathcal C$}
      \STATE Find $\pi^\star(p)$ and record
      $C(H,p,\pi^\star(p))$ with \textsc{Simulate}
    \ENDFOR
    \STATE Remove policies dominated under simulator cost
    \STATE Progressively measure $\Delta\mathrm{NLL}(p)$ and prune failures
    \RETURN the lowest-cost survivor with
    $\Delta\mathrm{NLL}(p)\leq\epsilon$
  \end{algorithmic}
\end{algorithm}

For every observed block $b$ and plane count $r$, the quantizer evaluates
legal exponent-gap templates $\mathcal T_r$. Using normalized maximum error
$h_{b,t}$ and MSE $m_{b,t}$, it selects the best template and aggregates its
excess error within each $R$-group:
\begin{equation}
\begin{aligned}
  e_{b,t}&=\alpha h_{b,t}+\beta m_{b,t},\qquad
  t^\star_{b,r}=\arg\min_{t\in\mathcal T_r}e_{b,t},\\
  E_{w,r}&=\frac{\sum_{b\in w}e_{b,t^\star_{b,r}}
  -\min_{r'}\sum_{b\in w}e_{b,t^\star_{b,r'}}}
  {|\mathcal B^{\rm obs}_{o(w)}|}.
\end{aligned}
\label{eq:dse-delta-elimination}
\end{equation}
Here $\mathcal B^{\rm obs}_{o(w)}$ contains the observed calibration blocks
in the QK or PV operator containing $w$. Operator-wise normalization prevents
QK's larger block count from overwhelming PV, while conditional template
selection removes $\boldsymbol\Delta$ from the global assignment.

The proposed system approximates hardware cost using active-plane work and
K/V storage:
\begin{equation}
  \widehat C_{w,r}=\lambda_{\rm plane}\,\bar n_w^{\rm visit}r+
  \lambda_{\rm footprint}\,\bar n_w^{\rm value}r,
  \label{eq:dse-hardware-proxy}
\end{equation}
where $\bar n_w^{\rm visit}$ and $\bar n_w^{\rm value}$ are normalized
expected block visits and stored values. For excess-error budget $\tau$,
the relaxed variable $x_{w,r}$ assigns plane count $r$ to group $w$:
\begin{equation}
\begin{aligned}
  \min_{\mathbf x}\quad &\sum_{w,r}\widehat C_{w,r}x_{w,r}
  \quad\text{s.t.}\quad \sum_{w,r}E_{w,r}x_{w,r}\leq\tau,\\[-2pt]
  &\sum_r x_{w,r}=1\ \ \forall w,\qquad
  0\leq x_{w,r}\leq1\ \ \forall w,r.
\end{aligned}
\label{eq:dse-proxy-relaxation}
\end{equation}

For each budget $\tau$, the proposed system solves the relaxation in
\eqnref{eq:dse-proxy-relaxation} and projects each group to its largest
fractional option. A first greedy pass repairs error-budget violations by
choosing the change with the highest error reduction per unit of added cost.
A second greedy pass uses the remaining error slack to maximize cost reduction
per unit of added error. Together with $t^\star_{b,R_w}$, the repaired
assignment defines a candidate policy.

As summarized in Algorithm~\ref{alg:multifidelity-dse}, sweeping $\tau$
produces a frontier $\mathcal C$ after removing duplicate and
surrogate-dominated policies. For each survivor, the formulation-based
simulator exhaustively evaluates the legal mappings in $\mathcal S(p)$ and
removes policies dominated under the deployment cost in
\eqnref{eq:dse-deployment-cost}. Progressively larger held-out subsets then
measure paired $\Delta\mathrm{NLL}$ against dense BF16 K/V, reject clear
failures early, and validate the remaining cost--quality frontier. The final
design is the lowest-cost survivor satisfying
$\Delta\mathrm{NLL}\leq\epsilon$.

\section{Evaluation}
\label{sec:evaluation}


\begin{figure*}[t]
    \centering
    \includegraphics[width=\textwidth]{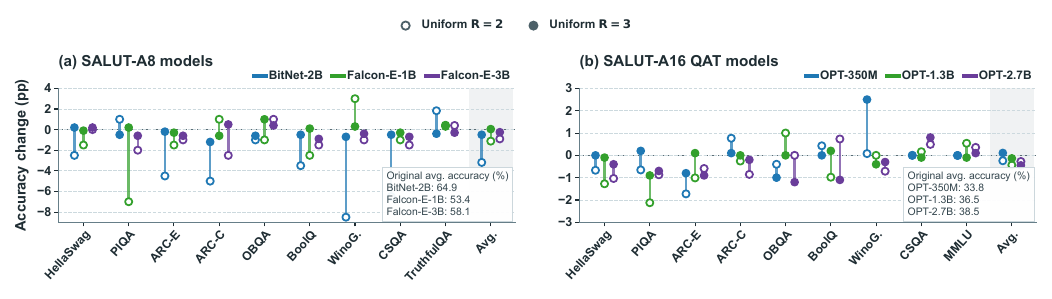}
    \caption{Full-test task-accuracy changes of uniform $R=2$ and $R=3$ relative to matched dense K/V. Each segment connects the two measured uniform settings. Insets list the nine-task average accuracy of each original checkpoint.}
    \label{fig:cross-model-mc9-r3}
\end{figure*}

We evaluate whether executable RSD K/V preserves model quality, improves hardware and inference efficiency, and enables useful quality--cost trade-offs. We then place the implementation results in the context of prior accelerators.

\subsection{Experimental Setup}

\textbf{Models.} We evaluate native ternary BitNet b1.58 2B4T and FalconE-1B/3B, together with QAT-converted ternary OPT-350M/1.3B/2.7B~\cite{BitNet158,BitNet2B4T,Spectra,qat-distillation}. The A8 configuration converts activation-side operands ($A/Q/P$) to INT8 before LUT construction, whereas the A16 configuration retains them in BF16. Both designs use ternary linear weights and signed-digit K/V. We use the A8 configuration for BitNet and Falcon-E and the A16 configuration for OPT, and compare quality with the same checkpoint using dense BF16 K/V.

\textbf{ASIC Evaluation.} To quantify the cost of provisioning separate linear and attention compute, we synthesize the proposed system and a throughput-matched Baseline in a TSMC 40-nm standard-cell library at 1.1\,V, with A8 at 500\,MHz and A16 at 250\,MHz. Both retain LUT execution for ternary linear and projection layers, while the Baseline uses dedicated dense MAC attention. In A8, the proposed system's DQEU quantizes $A/Q/P$ and encodes K/V, while the Baseline retains activation quantization and matches the fixed-$R=3$ QK/PV issue rate with INT8 MACs. In A16, the DQEU encodes only K/V for the proposed system and is absent from the BF16$\times$BF16-MAC Baseline. Dynamic power uses matched gate-level VCD windows in PrimeTime PX. The Baseline clock-gates its inactive LUT or MAC domain so that heterogeneous area is retained without charging avoidable switching.

\textbf{Simulator.} Our cycle-accurate model captures LUT construction, signed-digit plane execution, reductions, pipeline overheads, and data transfers for a given model, request, mixed-$R$ policy, hardware mapping, and bandwidth. It reports operator and request cycles, TTFT, TPOT, throughput, traffic, and K/V footprint. Unless noted otherwise, both systems use 16 cores, 250\,MHz, and 128\,GB/s. The equal-area A8 serving experiment is re-scored at the synthesized 500-MHz target. The DSE searches hardware-legal $(R,\boldsymbol{\Delta})$ policies and selects the fastest one that satisfies the held-out NLL constraint.

\textbf{Metrics.} Hardware speedup is relative to the matched Baseline, and operator efficiency is normalized throughput per area or active dynamic power. ASIC area and power are reported in $10^3\,\mu\mathrm{m}^2$ and mW. Absolute compute efficiency uses TOPS/mm$^2$ and TOPS/W, or TFLOPS/mm$^2$ and TFLOPS/W for BF16. TTFT is time to first token, and TPOT is the average interval between output tokens. We report external-memory traffic separately from packed K/V footprint and use it only as an indicator of memory-system energy~\cite{DataMovement}.

\subsection{Cross-Model Quantization Quality}

\figref{fig:cross-model-mc9-r3} compares uniform $R=2$ and $R=3$ with matched dense-K/V checkpoints. The inset in each panel reports the original checkpoint's average accuracy over the nine tasks. At the conservative $R=3$ setting, every model remains within 0.50~pp of dense K/V, showing that signed-digit K/V can preserve model quality across native and QAT-converted ternary models.

The larger model- and task-dependent variation at $R=2$ shows that lower plane counts reduce representation and execution costs at a nonuniform quality cost. A single global precision is therefore unnecessarily conservative for robust blocks and unsafe for sensitive ones. This observation motivates the block-wise mixed-$R$ policies evaluated in the DSE, where hardware benefits are accepted only after a held-out quality check.

\subsection{Hardware Evaluation}

\begin{table}[t]
  \centering
  \caption{Compute-tile area and active $QK^{\mathsf T}$ power at fixed $R=3$.}
  \label{tab:area_power_breakdown}

  \scriptsize
  \setlength{\tabcolsep}{3.2pt}
  \renewcommand{\arraystretch}{1.12}

  \resizebox{0.64\columnwidth}{!}{%
    \begin{tabular}{l r r r r r r}
      \toprule
      & \multicolumn{3}{c}{Area ($10^3\,\mu\mathrm{m}^2$)}
      & \multicolumn{3}{c}{Power (mW, $QK^{\mathsf T}$, $R=3$)} \\
      \cmidrule(lr){2-4}
      \cmidrule(l){5-7}
      System
      & \shortstack{LUT\\Core}
      & \shortstack{Encode\\Logic}
      & \shortstack{Dense\\Attention Array}
      & \shortstack{LUT\\Core}
      & \shortstack{Encode\\Logic}
      & \shortstack{Dense\\Attention Array} \\
      \midrule
      Proposed-A8
      & 58.67$^\ast$ & 1.32 & ---
      & 5.80 & 0.590 & --- \\
      Baseline-A8
      & 59.85$^\ast$ & --- & 38.62
      & 0.00 & --- & 12.70 \\
      Proposed-A16
      & 283.11 & 10.96 & ---
      & 23.60 & 0.374 & --- \\
      Baseline-A16
      & 289.72 & --- & 737.01
      & 0.00 & --- & 133.0 \\
      \bottomrule
    \end{tabular}%
  }

  \vspace{1pt}
  \parbox{\columnwidth}{\tiny
  $^\ast$ For Proposed-A8 and Baseline-A8, LUT-Core area includes the shared
  INT8 quantization logic. Encode Logic denotes the signed-digit encoder
  portion of the DQEU in A8 and the DQEU in A16.}
\end{table}

\begin{figure}[t]
    \centering
    \includegraphics[width=0.66\columnwidth]{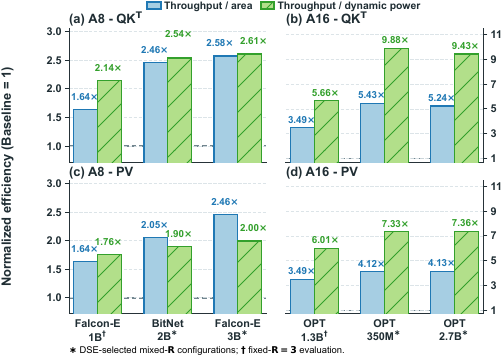}
    \caption{Normalized QK/PV throughput-per-area and throughput-per-power across
    evaluated models, relative to the heterogeneous Baseline with separate linear
    and attention engines. $\dagger$ denotes fixed-$R=3$ points; the remaining
    points use model-specific mixed-$R$ policies selected by the DSE, combining
    unified-LUT architectural gains with lower average plane counts.}
    \label{fig:model-specific-qk-pv-efficiency}
\end{figure}

At fixed $R=3$, each 64-value group takes three plane cycles. The
throughput-matched Baseline therefore uses a 22-lane dense attention array,
issuing 22/22/20 products with the two unused lanes clock-gated in the final
cycle. As shown in \tabref{tab:area_power_breakdown}, the encoder adds only
2.2\%/3.9\% of LUT-Core area in A8/A16, whereas the separate dense-attention
array adds 64.5\%/254.4\%. For $QK^{\mathsf T}$, the proposed system consumes
6.39/23.97\,mW, compared with 12.70/133.0\,mW for the Baseline attention
arrays.

In \figref{fig:model-specific-qk-pv-efficiency}, the inactive Baseline domain
is clock-gated during each PTPX window, while its full provisioned area is
included. For a mixed-$R$ policy with block fraction $f_r$, measured power
$P_r$, and $\bar R=\sum_r r f_r$, we use
\begin{equation}
  P_{\mathrm{mix}}
  =
  \frac{\sum_r f_r rP_r}{\bar R},
  \qquad
  T_{\mathrm{mix}}
  =
  \frac{3T_3}{\bar R},
  \label{eq:mixed-r-power}
\end{equation}
where $T_3$ is fixed-$R=3$ throughput.

The fixed-$R=3$ points isolate the architectural benefit because the proposed
system and the Baseline execute the same three-plane workload. Falcon-E-1B
reaches $1.64\times$/$2.14\times$ throughput-per-area/power for QK and
$1.64\times$/$1.76\times$ for PV on Proposed-A8, while OPT-1.3B reaches
$3.49\times$/$5.66\times$ and $3.49\times$/$6.01\times$ on Proposed-A16.
These gains therefore come from reusing the linear LUT and reduction datapath
for attention rather than provisioning a separate dense-attention array. The
larger A16 gains reflect the higher area and power cost avoided by replacing
the BF16 MAC array.

The mixed-$R$ points further benefit from DSE-selected lower plane counts,
which reduce execution cycles and switching activity without resizing the
compute substrate. Taking the arithmetic mean over QK and PV, Falcon-E-3B on
Proposed-A8 achieves $2.52\times$ throughput per area and $2.30\times$
throughput per power over the throughput-matched 40-nm Baseline. OPT-350M on
Proposed-A16 achieves $4.78\times$ and $8.61\times$, respectively. These
results show that architectural reuse and DSE-selected precision provide
complementary efficiency gains; the associated model-quality trade-offs are
quantified in Section~VI-E.

\subsection{Operator and Inference Efficiency}

\begin{figure}[t]
    \centering
    \includegraphics[width=0.56\columnwidth]{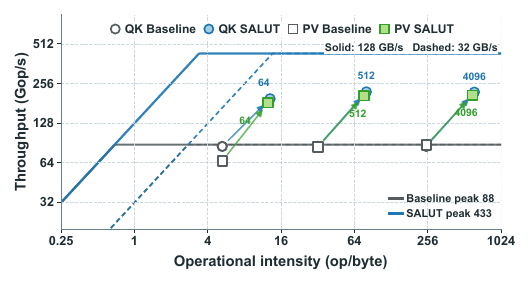}
    \caption{Area-matched QK/PV roofline for BitNet-2B under the selected mixed-$R$ policy across context lengths and memory bandwidths.}
    \label{fig:asic-a8-qk-pv-roofline}
\end{figure}

\figref{fig:asic-a8-qk-pv-roofline} uses the full compute-tile areas in \tabref{tab:area_power_breakdown}, 59.99 and 98.47~$10^3\,\mu\mathrm{m}^2$ for Proposed-A8 and Baseline-A8, and holds external bandwidth fixed under continuous equal-area normalization. Compressed K/V reads raise QK and PV operational intensity by $2.38$ to $2.50\times$, while the smaller unified tile raises attainable throughput under the same area budget. The realized points remain below the fixed-$R=1$ peak because the selected policy executes an average of more than one plane per value. Thus, the gain comes from both fewer transferred bytes and more useful attention work per unit area.

\begin{figure}[t]
    \centering
    \includegraphics[width=0.66\columnwidth]{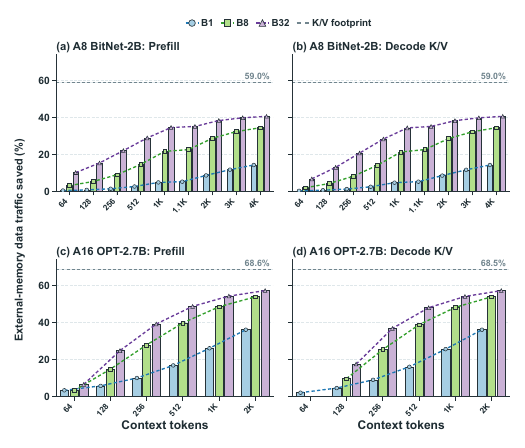}
    \caption{External-memory traffic and packed K/V footprint savings for BitNet-2B and OPT-2.7B across context length and batch size.}
    \label{fig:asic-a8-external-memory-sweep}
\end{figure}

\figref{fig:asic-a8-external-memory-sweep} measures total external-memory traffic, including transfers unaffected by K/V compression, as a proxy for memory-system energy~\cite{DataMovement}. Savings grow with context length and batch size because shared model traffic is increasingly amortized while per-sequence K/V traffic grows. They reach 40.5\% for BitNet-2B and 57.3\% for OPT-2.7B at the largest settings. For the OPT traffic sweep, we apply fixed-$R=3$ encoding across the modeled context range to isolate the context-length scaling of K/V traffic.

\begin{figure}[t]
    \centering
    \includegraphics[width=0.68\columnwidth]{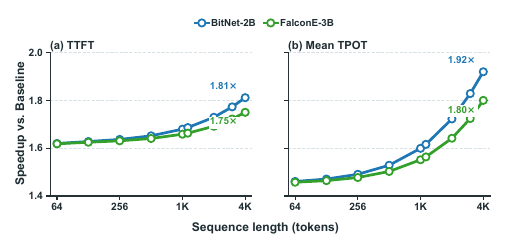}
    \caption{Estimated equal-area A8 request-level speedup at 500\,MHz and 128\,GB/s. Each point is the geometric mean over $B=1,8,32$, including scale and template metadata reads.}
    \label{fig:asic-a8-serving-impact}
\end{figure}

For request-level inference, integer provisioning replaces the continuous normalization with realizable tile counts. The area of 16 Baseline-A8 tiles accommodates 26 Proposed-A8 tiles after rounding down, using 99.0\% of the same budget. We re-score operator service at 500\,MHz and 128\,GB/s while retaining fixed pipeline overheads and charging K/V metadata reads. In \figref{fig:asic-a8-serving-impact}, the gains increase with context length as attention occupies more of each request. At 4K, TTFT improves by $1.81\times$ and $1.75\times$ for BitNet-2B and Falcon-E-3B, while mean TPOT improves by $1.92\times$ and $1.80\times$. TPOT benefits more strongly because every decode step revisits a growing cache.

\begin{table}[t]
  \centering
  \caption{Buffer-side overhead at 4,096 tokens and $B=1$. K/V and metadata are in MiB.}
  \label{tab:cache-metadata-vtail}
  \scriptsize
  \setlength{\tabcolsep}{2.8pt}
  \renewcommand{\arraystretch}{1.04}
  \resizebox{0.5\columnwidth}{!}{%
    \begin{tabular}{@{}lrrrr@{}}
      \toprule
      \textbf{Model} &
      \shortstack{\textbf{Packed}\\\textbf{K/V}} &
      \shortstack{\textbf{Metadata}\\\textbf{MiB (\% K/V)}} &
      \shortstack{\textbf{Total /}\\\textbf{BF16}} &
      \shortstack{\textbf{V-tail}\\\textbf{MiB (\% GB)}} \\
      \midrule
      BitNet-2B  & 61.66 & 7.25 (11.8\%) & 23.0\% & 1.91 (3.0\%) \\
      Falcon-E-3B & 26.15 & 3.09 (11.8\%) & 22.8\% & 0.81 (1.3\%) \\
      \bottomrule
    \end{tabular}%
  }
\end{table}

\tabref{tab:cache-metadata-vtail} charges one FP16 runtime scale and one 8-bit template ID per encoded block. Even after this charge, the cache remains below 23.0\% of dense BF16. The Open V-Block Region holds the V-tail state until a 32-token V block is finalized. This state comprises one authoritative BF16 tail block and two signed-digit shadow banks. Because it covers only the current partial block and is reused after finalization, it does not grow with sequence length and occupies only 1.3\% to 3.0\% of the shared 64-MiB Global Buffer at 4K.

\begin{figure}[t]
    \centering
    \includegraphics[width=0.56\columnwidth]{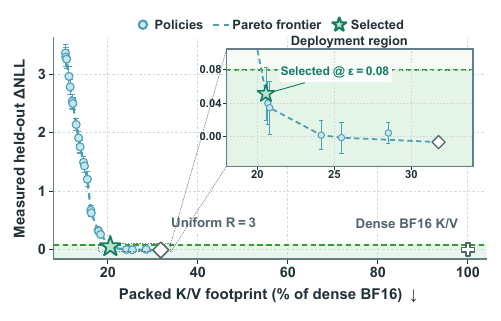}
    \caption{Measured BitNet-2B quality--K/V footprint frontier across
    25 hardware-legal block-32 policies at 4,096 tokens.}
    \label{fig:quality-footprint-pareto}
\end{figure}

\begin{figure*}[t]
    \centering
    \includegraphics[width=\textwidth]{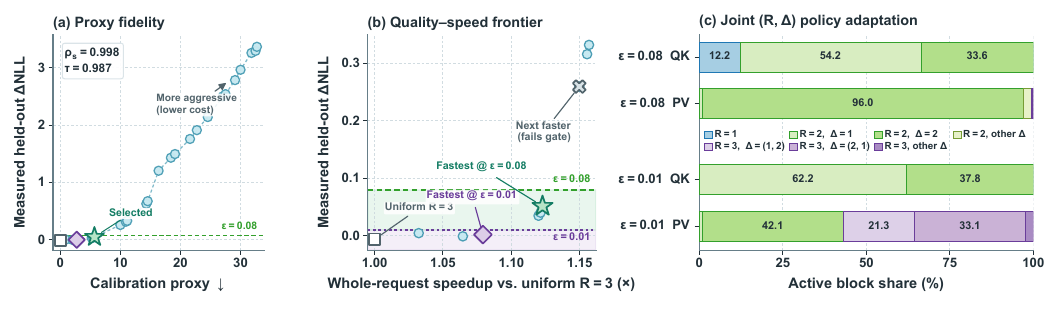}
    \caption{Measured BitNet-2B block-32 DSE at 4,096 tokens:
    (a) proxy fidelity, (b) quality--speed frontier, and
    (c) selected QK/PV policy composition.}
    \label{fig:dse-search-evaluation}
\end{figure*}

\subsection{Design Space Exploration}

\begin{table}[t]
  \centering
  \caption{Throughput benefit of two representative mixed-$R$ policies.}
  \label{tab:mixed-r-benefit}
  \scriptsize
  \setlength{\tabcolsep}{2.6pt}
  \renewcommand{\arraystretch}{1.04}
  \begin{tabular}{@{}lcc@{}}
    \toprule
    \textbf{Model} &
    \shortstack{\textbf{QK}\\
    $R=3\rightarrow\bar R$,\ $T/T_3$} &
    \shortstack{\textbf{PV}\\
    $R=3\rightarrow\bar R$,\ $T/T_3$} \\
    \midrule
    BitNet-2B (A8 configuration) &
    $3\rightarrow2.00$,\ $1.50\times$ &
    $3\rightarrow2.40$,\ $1.25\times$ \\
    OPT-350M (A16 configuration) &
    $3\rightarrow1.93$,\ $1.56\times$ &
    $3\rightarrow2.54$,\ $1.18\times$ \\
    \bottomrule
  \end{tabular}
\end{table}

We first quantify how selected plane counts translate into cycles. Since a
64-value group consumes one cycle per active plane, ideal throughput relative
to fixed $R=3$ scales as $3/\bar R$. As shown in
\tabref{tab:mixed-r-benefit}, the BitNet-2B policy improves QK/PV throughput
by $1.50\times$/$1.25\times$, while OPT-350M achieves
$1.56\times$/$1.18\times$ without resizing the core. The higher PV plane
counts lead to smaller PV speedups. Across the nine tasks, the selected
mixed-$R$ policies reduce mean accuracy by 0.72~pp for BitNet-2B, 1.86~pp
for Falcon-E-3B, 1.54~pp for OPT-350M, and 0.56~pp for OPT-2.7B relative to
their matched dense-K/V checkpoints.

We further examine BitNet-2B as a block-32 case study, separate from the
model-specific operating points in \tabref{tab:mixed-r-benefit}.
\figref{fig:quality-footprint-pareto} shows the capacity trade-off. Under
the primary $\epsilon=0.08$ NLL budget, the selected policy reduces packed
K/V footprint by 35.3\% relative to uniform $R=3$, achieves
$\Delta\mathrm{NLL}=0.0515$, and lies on the held-out quality--footprint
frontier.

\figref{fig:dse-search-evaluation} shows the selection process. LP projection and repair yield 25 hardware-feasible policies, the formulation-based simulator evaluates request latency, and paired inference measures $\Delta\mathrm{NLL}$ on eight held-out 4,096-token segments. The calibration proxy ranks candidate policies, while held-out NLL enforces the deployment constraint. Their Spearman correlation reaches $\rho_s=0.998$. Under $\epsilon=0.08$, the fastest feasible policy achieves a $1.123\times$ request-level speedup at $\Delta\mathrm{NLL}=0.0515$. The next faster policy reaches $1.150\times$ but is rejected at $\Delta\mathrm{NLL}=0.2594$. Tightening the budget to $\epsilon=0.01$ instead selects a $1.079\times$ policy at $\Delta\mathrm{NLL}=0.0021$, demonstrating that the quality gate controls the operating point. Finally, \figref{fig:dse-search-evaluation}(c) shows different QK and PV mixtures because $R$ determines plane cycles while $\boldsymbol{\Delta}$ changes the codebook at fixed hardware cost. All serving sweeps use the selected primary policy without per-request retuning.

\subsection{Comparison with Related Accelerators}

Finally, we compare the proposed system with representative LLM accelerators at two levels. The same-node and same-library Baselines in
\tabref{tab:area_power_breakdown} and
\figref{fig:model-specific-qk-pv-efficiency} provide controlled architectural
comparisons, while \tabref{tab:compute-efficiency} places the proposed system
in the context of prior implementations using each design's reported process
and computation boundary.

For Proposed-A8 Projection, the complete INT8$\times$W1.58 datapath achieves
5.38\,TOPS/W, compared with 3.25\,TOPS/W for BitMoD and 5.12\,TOPS/W for
Tender. For runtime attention, Proposed-A16 achieves 0.69\,TFLOPS/W for
QK/PV, while FIGLUT reports 0.47\,TFLOPS/W. Thus, the proposed system provides
comparable compute efficiency despite using a more conservative 40-nm, 1.1-V
process rather than the 28-nm processes of these designs. We additionally
report the 29.01\,TOPS/W Proposed-A8 LUT-array-only result at the computation
boundary used by LUT Tensor Core. The QK/PV results use quality-selected
average plane counts of $\bar R=1.92$ for A8 and $\bar R=2.13$ for A16,
demonstrating that the shared LUT substrate remains effective for dynamic K/V
execution.

\begin{table}[t]
  \centering
  \caption{Compute efficiency comparison.}
  \label{tab:compute-efficiency}
  \scriptsize
  \setlength{\tabcolsep}{0.4pt}
  \renewcommand{\arraystretch}{0.98}
  \begin{tabular}{@{}>{\raggedright\arraybackslash}p{0.18\columnwidth}
  >{\raggedright\arraybackslash}p{0.26\columnwidth}
  >{\centering\arraybackslash}p{0.15\columnwidth}
  >{\centering\arraybackslash}p{0.17\columnwidth}
  >{\centering\arraybackslash}p{0.17\columnwidth}@{}}
    \toprule
    \textbf{Work} &
    \textbf{Compute boundary} &
    \textbf{Node / V / MHz} &
    \begin{tabular}[t]{@{}c@{}}
      \textbf{Area eff.}\\[-1pt]
      {\tiny (TOPS/mm$^2$)}
    \end{tabular} &
    \begin{tabular}[t]{@{}c@{}}
      \textbf{Energy eff.}\\[-1pt]
      {\tiny (TOPS/W)}
    \end{tabular} \\
    \midrule

    LUT Tensor Core~\cite{LUTTensorCore} &
    INT1$\times$INT8 LUT PE without PSUM &
    28\,nm / 0.9\,V / 1000 &
    -- & 63.78 \\

    BitMoD~\cite{BitMoD} &
    FP16$\times$b1 bit-serial tile &
    28\,nm / -- / 1000 &
    1.29$^\ddagger$ & 3.25$^\ddagger$ \\

    Tender~\cite{Tender} &
    INT4 systolic array &
    28\,nm / -- / 1000 &
    2.06$^\ast$ & 5.12$^\ast$ \\

    FIGLUT~\cite{FIGLUT} &
    FP--INT LUT GEMM &
    28\,nm / -- / 100 &
    -- &
    \begin{tabular}[t]{@{}c@{}}
      0.47\\[-1pt]
      {\tiny (TFLOPS/W)}
    \end{tabular} \\

    Slim-Llama~\cite{SlimLlama} &
    Binary/ternary LLM processor &
    28\,nm / 0.58--1.0\,V / 25--200 &
    0.243 & 60.0 \\

    \midrule

    \rowcolor[rgb]{0.902,0.965,0.933}
    \textbf{Proposed-A8} &
    Projection, INT8$\times$W1.58 &
    40\,nm / 1.1\,V / 500 &
    \begin{tabular}[t]{@{}c@{}}
      1.07\\[-1pt]
      {\tiny (3.74$^\dagger$)}
    \end{tabular} &
    \begin{tabular}[t]{@{}c@{}}
      5.38\\[-1pt]
      {\tiny (29.01$^\dagger$)}
    \end{tabular} \\

    \rowcolor[rgb]{0.902,0.965,0.933}
    \textbf{Proposed-A8} &
    QK/PV, INT8$\times$RSD
    (Avg.\ $\bar R=1.92$) &
    40\,nm / 1.1\,V / 500 &
    0.55 & 3.95 \\

    \rowcolor[rgb]{0.902,0.965,0.933}
    \textbf{Proposed-A16} &
    Projection, BF16$\times$W1.58 &
    40\,nm / 1.1\,V / 250 &
    \begin{tabular}[t]{@{}c@{}}
      0.109\\[-1pt]
      {\tiny (TFLOPS/mm$^2$)}
    \end{tabular} &
    \begin{tabular}[t]{@{}c@{}}
      1.59\\[-1pt]
      {\tiny (TFLOPS/W)}
    \end{tabular} \\

    \rowcolor[rgb]{0.902,0.965,0.933}
    \textbf{Proposed-A16} &
    QK/PV, BF16$\times$RSD
    (Avg.\ $\bar R=2.13$) &
    40\,nm / 1.1\,V / 250 &
    \begin{tabular}[t]{@{}c@{}}
      0.051\\[-1pt]
      {\tiny (TFLOPS/mm$^2$)}
    \end{tabular} &
    \begin{tabular}[t]{@{}c@{}}
      0.69\\[-1pt]
      {\tiny (TFLOPS/W)}
    \end{tabular} \\

    \bottomrule
  \end{tabular}

  \vspace{1pt}
  \parbox{\columnwidth}{\tiny
  $^\dagger$ Parenthesized A8 Projection values use the LUT-array-only boundary
  of LUT Tensor Core~\cite{LUTTensorCore}. Unparenthesized values include the
  complete datapath.
  $^\ast$ Tender retains its reported INT4 accounting (two operations per MAC).
  $^\ddagger$ BitMoD retains its reported bit-serial accounting.}
\end{table}
\section{Conclusion}

This paper presents a quantization--architecture co-design that resolves the
representation and layout mismatch between ternary linear LUT execution and
dynamic K/V attention. By treating ternary weights as the single-plane
($R=1$) case of a restricted signed-digit representation, the proposed system
encodes runtime K/V into packed ternary planes that remain directly executable
through the same activation-derived LUT interface without reconstructing dense
cache entries. To satisfy the different reduction structures of $QK^\top$ and
$PV$, it uses operator-aligned K/V layouts and a bounded V-tail mechanism that
maintains scale-consistent encoding during incremental cache updates. Its
unified multi-stream architecture shares LUT construction, lookup, reduction,
and post-processing across linear, $QK^\top$, and $PV$, while a hardware-aware
DSE coordinates signed-digit precision, block geometry, and stream mapping
under quality and implementation constraints.

Overall, this work establishes K/V cache quantization as a directly executable
computation format for LUT-based attention. This capability extends ternary
LUT computation from static linear projections to dynamic attention on a
shared compute substrate.
\section{Acknowledgements}

The authors would like to express deep appreciation to LuxiTech Co. Ltd. for the essential development and experimental resources support. Furthermore, this work was partially funded by the Huazhong University of Science and Technology Cross Research Support Program (2024JCYJ036) and the Wuhan Municipal Key Science and Technology Program (2022010402020045), and was partially conducted by ACCESS – AI Chip Center for Emerging Smart Systems, supported by the InnoHK initiative of the Innovation and Technology Commission of the Hong Kong Special Administrative Region Government.

\bibliographystyle{unsrt}  
\bibliography{refs}

\end{document}